\documentclass[amssymb,twocolumn,
reprint,aps,pra,floatfix,showpacs]{revtex4}
\usepackage{graphicx}

\newcommand{\intensity}[2]{$#1\times10^{#2}\mbox{ Wcm}^{-2}$}
\newcommand{\mz}{$M=0$ }
\newcommand{\mo}{$M=1$ }

\begin{document}
\title{Enhanced Harmonic Generation from $M=1$ aligned Ar$^+$}
\author{A. C. Brown and H. W. van der Hart}
\affiliation{Centre for Theoretical Atomic, Molecular and Optical Physics,
Queen's University Belfast, Belfast, BT7 1NN, UK.}
\date{}

\begin{abstract} We investigate harmonic generation (HG) from ground-state
  Ar$^+$ aligned with \mo at a laser wavelength of 390-nm and intensity of
  \intensity{4}{14}. Using time-dependent R-matrix theory, we find that an initial
  state with magnetic quantum number \mo provides a 4-fold increase in harmonic
  yield over $M=0$.  HG arises primarily from channels associated with the $^3P^e$
  threshold of Ar$^{2+}$, in contrast with \mz for which channels associated with
  the excited, $^1D^e$ threshold dominate HG. Multichannel and
  multielectron interferences lead to a more marked suppression of HG for \mo
  than $M=0$.  \end{abstract}

\maketitle

\section{Introduction} 

Recent advances in laser technology have facilitated the experimental
analysis of atomic structure and electron dynamics in more detail than ever before
\cite{corkum_review}. The process of harmonic generation (HG) has been
central to these advances both as a basis for ultrashort and high energy light
pulses \cite{67as_pulse,efficient_kev_hhg}, and as a tool for
elucidating ultrafast dynamics \cite{real_time_tunnelling,
real_time_valence_motion, chemical_reaction_hhg}.
These dynamics can be strongly influenced by the interaction of
channels associated with different ionization thresholds
\cite{multiple_orbitals_N2,multielectron_molecules,multichannel_ionization,multielectron_atoms}.
Hence, the development of methods capable of investigating the
influence of channel coupling on HG is of strong current interest.

Traditionally, the HG process is thought of as a single electron process
\cite{corkum1993}, and a variety of theoretical methods for addressing the
process have been based on this approach \cite{lewenstein,
sae_hhg_kulander, sae_hhg_ivanov}. The `three-step', or recollision, model of HG outlined in
\cite{corkum1993} describes a laser-ejected electron being driven by the field
before recolliding with its parent ion. Upon recollision the electron can be
recaptured, emitting its energy in the form of an odd harmonic of the driving
laser frequency. Despite the success of this approach,
recent research has uncovered more complex multielectron and multichannel
influences in HG
\cite{multiple_orbitals_N2,multielectron_molecules,multielectron_atoms}.
Thus there is a need for quality theoretical data to direct experimental
attosecond science, and to enhance our
understanding of these fundamental physical processes.

It is for these reasons that we have developed time-dependent R-matrix (TDRM)
theory \cite{tdrm}-- a
fully non-perturbative, {\it ab initio } method for the description of general
multielectron atoms in short, intense laser pulses. TDRM has been used
successfully to elucidate the collective electron-response to laser light in
C$^+$, complete with a detailed comparison of the dynamics arising from
different initial magnetic orientations (\mz or $M=1$) \cite{collect_c+1,
collect_c+_m1}. More recently the method has been extended to address HG
\cite{brown_helium}, uncovering multielectron interference leading to resonant
enhancement of HG from argon \cite{brown_prl} and multichannel
interference leading to suppression of HG in Ar$^+$ \cite{brown_ar+}. It is this
Ar$^+$ system which occupies us in this article:
accurate treatment of the electron emission channels associated with the
different, closely-spaced $3s^23p^4$ thresholds requires
a method capable of describing both
multielectron and multichannel effects.

The results contained in this article extend the work presented in
\cite{brown_ar+} to investigate the effect of the initial magnetic quantum
number, $M$, on HG. Although ground-state noble gas targets, predominantly used in HG
experiments, can only have $M=0$, it has been suggested that the
highest harmonics produced from Ar arise from ionized species generated
during the laser-Ar interaction \cite{argon+_zepf,
argon+_gibson,argon+_wahlstrom}.  We have
previously demonstrated \cite{brown_ar+} that despite the general assumption
that a higher ionization potential leads to a lower harmonic yield
\cite{hhg_factors}, the yield from Ar$^+$ with \mz actually exceeds that of He, even
though Ar$^+$ is more strongly bound. By analyzing the yield from \mo
we can improve our understanding of how
atomic structure affects HG.

The magnetic quantum number, $M$ determines which radiative transitions are
permitted according to the dipole selection rules. The first difference for $M\ne
0$ is that $\Delta L = 0$ transitions are allowed. We can thus have transitions
from $^2P^o$ states to $^2P^e$ and $^2D^e$ states. Further changes arise as for
\mz the five $3p$ electrons occupy $m=\{-1,-1,0,1,1\}$, whereas for
$M=1$ they occupy $m=\{-1,0,0,1,1\}$. Thus, \mo has two $m=0$ electrons in the outer shell while \mz has
only one; it is these $m=0$ electrons which dominate the response to
the laser field. Furthermore, for $M=1$, emission of an $m=0$ electron leaves $3p^4$ in
$\{-1,0,1,1\}$, which can form both triplet and singlet spin. Conversely, for
$M=0$, emission of an $m=0$ electron leaves $3p^4$ in $\{-1,-1,1,1\}$, which
couples only to singlet spin. Hence emission of an $m=0$ electron leaving
Ar$^{2+}$ in its $^3P^e$ ground state is possible only for $M=1$.
This change may lead to enhanced
ionization: application of the ADK formula \cite{ADK} suggests that the
ionization rate could be increased by a factor of 5~in the \mo case.
Finally, for $M=0$, ionization can proceed via
intermediate excitation of the $3s3p^6$ $^2S$ bound-state. However, specifying \mo precludes the
system having $S$ symmetry. It is therefore of interest to study how a change in
$M$ affects HG for Ar$^+$.


\section{Theory} 

A full account of the TDRM method for $M\ne 0$ can be found in
\cite{tdrm,collect_c+_m1}, and the extension of TDRM
to HG in \cite{brown_helium}. Here we give only a brief overview. TDRM employs
the standard R-matrix partition of space such that within a particular radius of
the nucleus all interactions between electrons are fully described. Should a photoionized
electron pass beyond the boundary of this region it becomes spatially isolated
from the residual ion. Electron-exchange effects can then be neglected, and
the electron moves only in the long-range potential of the residual ion and the
laser field. Then, starting from a field-free solution, the time-dependent
Schr\"odinger equation can be solved, using a Crank-Nicolson scheme to propagate
the wavefunction in time.

HG arises from the laser-driven oscillation of the atomic
dipole. The harmonic spectrum is proportional to the charge's acceleration, but
can be calculated via either the dipole length or velocity.
In TDRM theory, both methods give the same result to a high level of accuracy
for He
\cite{brown_helium}. The spectra shown here are calculated from the expectation value
of the dipole length operator:
\begin{equation}
  \label{eq:dipole_length}
  \mathbf{d}(t)\propto \left< \Psi (t) |
  \mathbf{z}  | \Psi (t) \right>, \end{equation}
where $\mathbf{z}$ is the total position operator along the laser polarization
axis. The harmonic spectrum is then proportional to
\begin{equation}
  \label{prop_dip}
\omega^4|\mathbf{d}(\omega)|^2,
\end{equation}
where $\omega$ is the laser frequency and $\mathbf{d}(\omega)$ is the Fourier
transform of $\mathbf{d}(t)$. With this approach we can extract the
single-atom/ion response of the system. We do not consider macroscopic effects,
but rather treat the single-atom in a level of detail not afforded by any other method.


\section{Computational setup}

The Ar$^+$ targets used in this article are as described in \cite{brown_ar+}. We
calculate the $3s^23p^4$ and $3s3p^5$ eigenstates of Ar$^{2+}$ via
configuration-interaction calculations, and describe Ar$^+$ as one of these
states plus an additional electron. This method allows us to vary systematically
the atomic structure contained in the calculations, and assess the
effects of including various thresholds. We employ several models in our
calculations. The five-state description comprises all five Ar$^{2+}$
thresholds. The three-state model includes only the $3s^23p^4$ thresholds. We
also use models which comprise individual $3s^23p^4$ $^3P^e$, $^1D^e$ or $^1S^e$
thresholds with or without the $3s3p^5$ thresholds. The initial
state is the Ar$^+$ $3s^23p^5$ ground-state with $M=1$.

As described above, exchange effects are included within a radius of
15~a.u. of the nucleus.
We use $60$ B-splines per angular momentum
for the description of the continuum orbitals. The Ar$^+$ basis then comprises
all allowed combinations of these continuum orbitals with the Ar$^{2+}$ states
up to a maximum angular momentum of $L=19$. The outer region is divided
into sectors of 2~a.u. each containing 35 9th order $B$-splines. The time-step
for the wavefunction propagation is 0.1~a.u. We use a 390-nm, \intensity{4}{14}
spatially homogeneous and linearly polarized laser pulse comprising a 3-cycle
$\sin^2$ ramp-on/ramp-off, and 2 cycles at peak
intensity.


\section{Results} 
\begin{figure}[t]
	\centering
	\includegraphics[width=8.6cm]{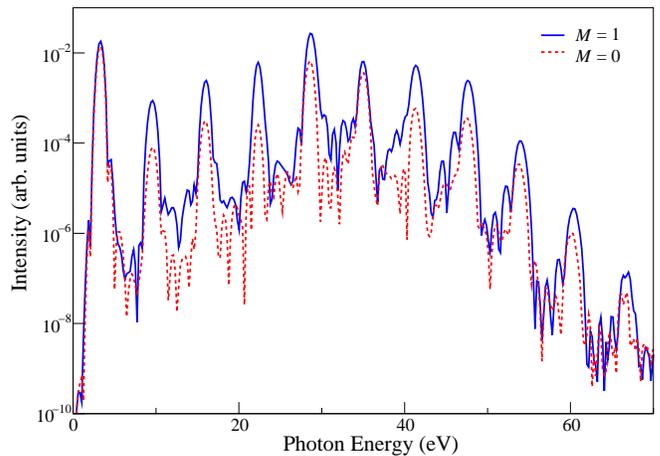}
	\caption{ (Color online) The harmonic spectra produced by Ar$^+$ in the three-state
	  model (see text) for a 390-nm,
	  \intensity{4}{14} laser pulse, from both the \mz (red dashed line) and \mo
	  (blue, solid line) initial alignment. \mz results previously published in
	  \cite{brown_ar+}.  \label{3targ_m1vsm0}}
\end{figure}

Figure \ref{3targ_m1vsm0} shows the harmonic spectrum produced by the three-state Ar$^+$
target in both the \mo and \mz initial alignments. The
ionization yield from \mo is four times larger than that from \mz,
and we might naively assume that the \mo harmonic yield
should increase correspondingly. While an increase is evident, the
enhancement of the \mo harmonic yield is not trivially proportional
to the increase in the level of ionization.
In
general, the \mz harmonic peak values are between 10\% and 30\% of the
equivalent \mo peaks, except in the 1st, 7th and 11th harmonics, where the \mz peaks
are 77\%, 4\% and 55\% of the \mo values respectively. On
average the \mo yield is four times larger than for $M=0$,
which is in line with the increase in ionization.

\begin{figure}[t]
	\centering
	\includegraphics[width=8.6cm]{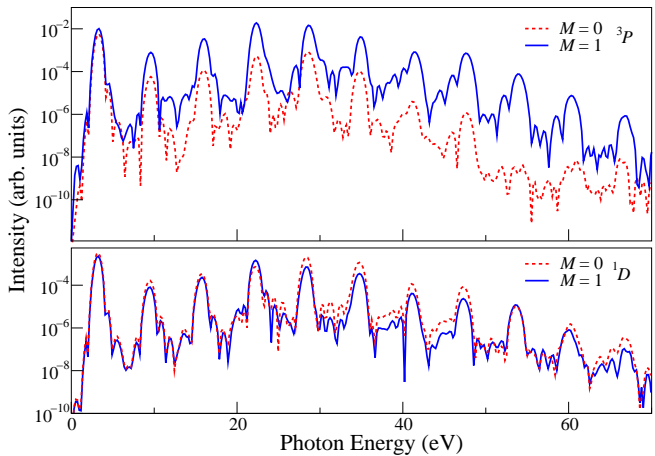}
	\caption{ (Color online) The harmonic spectra produced from the individual
	  $^3P$ (upper) and $^1D$ (lower) threshold description of Ar$^+$ with both $3s3p^5$
	  thresholds also active, for \mz (red, dashes) and \mo (blue line). \mz
	  results from \cite{brown_ar+}.
	  \label{singcomp} }
\end{figure}

To assess the physics underlying the changes between the \mz and \mo harmonic
spectra, we analyze the changes effected by the description of the atomic
structure. These changes are shown most dramatically by considering the harmonic
response of Ar$^+$ when only a single state of the $3s^23p^4$ configuration is
included. Figure \ref{singcomp} compares the spectra for \mo
and \mz calculations in which the individual $3s^23p^4$ $^3P^e$ and $^1D^e$
thresholds, and both $3s3p^5$ thresholds are included. The yield obtained in the
$^3P^e$ threshold calculation is dramatically enhanced for $M=1$:  the harmonic
peaks below the ionization energy are increased by a factor of 30, while those
above are increased by several orders of magnitude. This is mainly due to a 10-fold increase in the ionization yield associated with the
$^3P^e$ threshold for $M=1$. The increased harmonic yield associated with the
$^3P^e$ threshold does not substantially affect the $^1D^e$ spectrum, which is
reduced by only 30\% on average. Ionization towards the $^1D$
threshold also shows a 30\% decrease from $M=0$ to $M=1$, which agrees
with naive statistical calculations of the ionization probabilities.

For \mz it was found that the dominant contribution to
the harmonic yield was from channels connected to the first excited
threshold- $3s^23p^4$ $^1D^e$.  However, for \mo harmonic radiation stems primarily from the
Ar$^{2+}$ ground-state threshold- $3s^23p^4$ $^3P^e$.
For $M=1$, $m=0$ electron emission
channels associated with the $^3P^e$ state of Ar$^{2+}$ are available,
but they are not for $M=0$. A secondary factor is the presence of two $m=0$
electrons for $M=1$ compared to only one for $M=0$. Consequently, Fig.
\ref{singcomp} shows a dramatic increase in the
efficiency of HG for $M=1$ compared to $M=0$ in the $^3P^e$ calculation.

Figure \ref{singcomp} shows the dramatic effect of the presence of $m=0$
emission channels on the physics of HG.
The $^3P^e$, $M=0$, spectrum shows a strong harmonic response up to the ionization
threshold at 27 eV, but then falls off abruptly above this energy. The \mo
spectrum shows the more characteristic plateau of harmonics, extending to 45~eV,
approximately equal to the expected cut-off energy \cite{cutoff_law}. Thus the
\mo spectrum shows features consistent with the predictions of the recollision
model, whereas the \mz spectrum does not. Hence, a small change in the magnetic
quantum number can affect dynamics in a fundamental manner. In this case, the
interaction between the initial state and the continuum associated with the
$^3P^e$ threshold, critical for HG, has been altered on a fundamental level.

Figure \ref{singcomp} shows a slight increase in the $^1D^e$ yield for $M=0$
which may be due to the coupling of the $3s3p^6$ $^2S$ and $3s^23p^4nd$ $^2S$
states. For $M=1$, the $^2S$ states cannot be excited and hence the $3s3p^6$
state cannot act as an intermediate resonance.  From the separate $^1D^e$ and
$^3P^e$ spectra, Fig. \ref{singcomp} implies that the HG mechanism has changed
completely with an apparently small change in the atomic structure of the
target, and suggests that the three-step model may require extensions for the
description of HG in general systems.

The total harmonic yield cannot be approximated by that of a single threshold,
nor by trivially summing the individual contributions. Adding the $^3P^e$ and
$^1D^e$ spectra overestimates the total yield by as much as a factor of 20 for the
5th and 7th harmonics. Interferences between the $^3P^e$ and $^1D^e$ channels lead
to suppression of the harmonic yield.  The yield from the $^1S$ threshold is
four orders and two orders of magnitude smaller than that from the $^3P^e$ or
$^1D^e$ thresholds respectively. Multichannel effects involving the $^1S$
threshold do not have as dramatic an influence as $^3P^e$, $^1D^e$ interferences, but
still lead to a 50\% reduction in several harmonic peaks.

\begin{figure}[t]
	\centering
	\includegraphics[width=8.6cm]{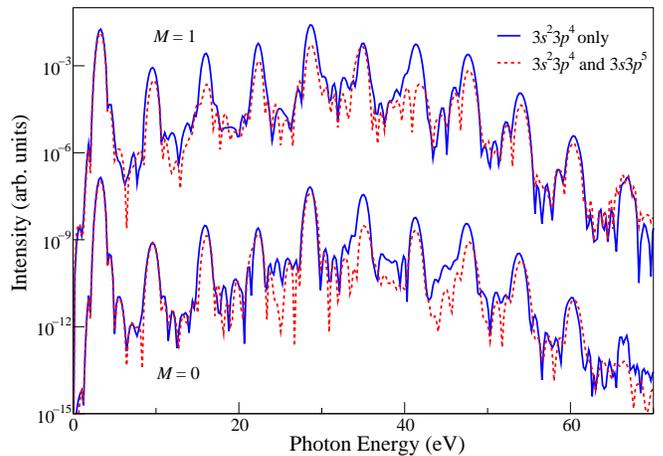}
	\caption{ (Color online) The harmonic spectra produced by the three- and
	  five-state models of Ar$^+$ for \mz and $M=1$. The \mz spectra have been
	  offset by a factor of $10^5$, and were previously published in
	  \cite{brown_ar+}.
	  \label{5targcomp}}
\end{figure}

In addition to these multichannel effects, the system also exhibits
interference between the response of $3s$ and $3p$ electrons.  Figure \ref{5targcomp} shows the effect of including
the $3s3p^5$ thresholds on the overall harmonic yield for both \mz and $M=1$.
There are noticeable differences in the way these interferences affect HG for
the different values of $M$.
Including the $3s3p^5$ thresholds in the \mo calculation leads to a factor 24
reduction in the 13th harmonic peak, the energy of which coincides with the $3s3p^5$ $^3P^o$
threshold. There is an order of magnitude reduction at the 5th harmonic and the
3rd, 7th, 9th and 15th peaks are all reduced by between three and five times.
In the \mz spectrum the largest effect of this inclusion is an order of magnitude decrease
in the 11th harmonic peak, with significant reductions also observed at the 13th
and 15th, and a 50\% suppression of the 5th harmonic.

The main difference between the spectra is thus that the harmonics below the
ionization threshold (3rd-9th) are more significantly reduced for \mo than for
$M=0$.  This can be understood in terms of field-driven, single-electron
transitions between $3s^23p^4 n\ell$ and $3s3p^5 n\ell$ Rydberg states.  Because
of the increased importance of the $^3P^e$ threshold in the \mo calculation
a significant population in $^3P^en\ell$ Rydberg states is expected. These
couple to the $^3P^on\ell$ states via excitation of a $3s$ electron to $3p$.
Such transfer may affect the coherent phase of the $3p$ electrons, thus
suppressing harmonic radiation which depends strongly on this coherence. While
such transitions are also expected for $M=0$, the low yield from the $^3P^e$
threshold reduces the appearance of the effect.  Single-electron transitions
between the $3s^23p^4 (^1D^e) n\ell$ or ($^1S^e) n\ell$ and $3s3p^5 (^1P^o) n\ell$
states are also important but the larger energy gap -- 7 photons in our
calculations versus 5 photons for the $^3P^e$ $\rightarrow$ $^3P^o$
transitions -- may reduce the impact on the low-energy harmonics for $M=0$. We note,
however, that the effect may be stronger in practice: using experimental
energies the energy gap corresponds approximately to a four-photon transition
in both cases.

Inclusion of the $3s3p^5$ thresholds has a noticeable effect on the 11th to 15th
harmonics, as shown in Fig. \ref{5targcomp}. For $M=1$, we see a reduction in
the intensity of the 13th harmonic by 1.5 orders of magnitude. This photon
energy overlaps the Rydberg series leading up to the $3s3p^5$ $^3P^o$ threshold.
With the dominant channels for HG being those connected with $^3P^e$, $3s-3p$
transitions connect the important $3s^23p^4$ $(^3P^e)$ $n\ell$ states
with $3s3p^5$ $(^3P^o)$ $n\ell$ states. This coupling could strongly influence
the 13th harmonic.  Similarly, for $M=0$, the dominant $^1D^e$ threshold can
couple to the $^1P^o$ threshold, leading to suppression of HG in the $3s3p^5$
$(^1P^o)$ $n\ell$ energy range-- the 15th harmonic in our calculation. This
interference is also responsible for the decrease in the 11th harmonic for
$M=0$, but the dominance of the $^3P^e$ channel means that this interference has
little effect for $M=1$.


\section{Discussion} 

The noticeable increase in the harmonic yield for $M=1$ relative to $M=0$ is
of significant importance for both theoretical and experimental treatment of
HG. From a theoretical standpoint, it is clear that apparently small details of
atomic structure can have a large bearing on the single-atom response to laser
light. It will be of great interest to see how experiment will bear out these
findings, as methods of extracting the single atom response from a
macroscopic picture will need to be developed.

Experimentally, the highest harmonics from Ar have been attributed to HG from
Ar$^+$ \cite{argon+_zepf,argon+_gibson}. This
constitutes a sequential process wherein Ar first ionizes and then undergoes HG.
Although ionization by a linearly polarized laser field will leave Ar$^+$
predominantly in the \mz state, the ion will subsequently evolve via the
spin-orbit interaction.  After 12~fs a significant \mo
population will arise. This transfer to \mo can then lead to a significant
enhancement of the harmonic response of Ar$^+$. It would be of interest to
investigate experimentally whether there are fundamental differences in HG from
Ar$^+$ or Ar for different pulse lengths.

Given the significant enhancement of the harmonic yield demonstrated here, and
the drive to improve the conversion efficiency of HG
\cite{improved_efficiency_hhg}, it is evident that the development of a
theoretical method incorporating the spin-orbit effect is of pressing
importance. Such a method will yield theoretical data invaluable in directing
experiment, and permit a more detailed analysis of the complex
dynamics of ultrafast processes.  To our knowledge, no method currently exists
which comprises the spin-orbit, multielectron and multichannel interactions, but
work is ongoing to extend TDRM theory to this end.


\section{Conclusion} 
We have used the TDRM method to describe HG from Ar$^+$ in both the
\mz and \mo initial alignments. The harmonic yield from \mo is enhanced by 4
times, on average, over the \mz yield, although there is a non-uniform increase
across the spectrum.  There is a noticeable difference in the way that
atomic structure affects the harmonic yield between \mz and $M=1$. The dominant
channels for HG are those connected to the ground ($^3P^e$) state of the Ar$^{2+}$
ion, which is at variance with the \mz case where the first excited ($^1D^e$)
state dominates. We have shown that multielectron and multichannel interferences
are observed for $M=1$, despite the apparent dominance of the $^3P^e$ threshold.  In
the \mo case these interferences lead to a more pronounced suppression of the
harmonic yield than for $M=0$.

While these calculations have been carried out using a wavelength of 390-nm we
expect that the conclusions are applicable for longer wavelengths. The energy
gap between the three lowest ionization thresholds in Ar$^+$ is approximately
equal to that of an 800-nm photon, and even if the channels associated with
higher lying thresholds were disfavoured at longer wavelengths, we have shown
that even channels with a very small contribution to the total ionization yield
can have a large impact on the total harmonic spectrum. The significant
influence of multielectron and multichannel effects on the process demonstrates
that a multielectron code, such as TDRM theory, or R-matrix incorporating time
(RMT) \cite{RMT}, is essential for the accurate description of HG.


 \section{Acknowledgements} 
ACB acknowledges support from DEL (NI). HWH is supported by EPSRC under
grant number G/055416/1.

\bibliography{mybib}
\end{document}